\documentclass{aastex631}
\usepackage{CJK}

\submitjournal{AJ}

\shortauthors{Wang et al.}

\begin{document}
\begin{CJK*}{UTF8}{gbsn}

\title{FX UMa: A New Heartbeat Binary System with Linear and Non-linear Tidal Oscillations and $\delta$ Sct Pulsations} 

\correspondingauthor{Kun Wang}
\email{kwang@cwnu.edu.cn}

\author[0000-0002-5745-827X]{Kun Wang(王坤)}
\affiliation{School of Physics and Astronomy, China West Normal University, Nanchong 637009, People's Republic of China}

\author{Anbing Ren(任安炳)}
\affiliation{School of Physics and Astronomy, China West Normal University, Nanchong 637009, People's Republic of China}

\author{Mads Fredslund Andersen}
\affiliation{Stellar Astrophysics Centre, Aarhus University, DK-8000 Aarhus C, Denmark}

\author[0000-0002-8736-1639]{Frank Grundahl}
\affiliation{Stellar Astrophysics Centre, Aarhus University, DK-8000 Aarhus C, Denmark}

\author{Tao Chen(陈涛)}
\affiliation{School of Physics and Astronomy, China West Normal University, Nanchong 637009, People's Republic of China}

\author{Pere L.~Pall\'e}
\affiliation{Instituto de Astrofísica de Canarias. E-38205 La Laguna, Tenerife, Spain }
\affiliation{Universidad de La Laguna (ULL), Departamento de Astrofísica, E-38206 La Laguna, Tenerife, Spain}

\begin{abstract}
We present a detailed analysis of an eclipsing double-lined binary FX UMa 
based on $\it TESS$ photometry and newly acquired spectroscopic observations.
The radial velocities and atmospheric parameters for each component star are obtained from the $\it SONG$ high-resolution spectra.
Combined with the radial-velocity measurements, our light-curve modeling 
yields absolute masses and radii of the two components.
The Fourier amplitude spectrum of the residual light curve
reveals a total of 103 frequencies with signal-to-noise ratio (S/N) $\geq$ 4, 
including 12 independent frequencies, 
17 multiples of the orbital frequency ($Nf_{orb}$), and 74 combination frequencies.
Ten $Nf_{orb}$ peaks with S/N $>$ 10 have very high amplitudes and are likely due to tidally excited oscillations (TEOs). 
The remaining $Nf_{orb}$ peaks (4 $\leq$ S/N $\leq$ 10) may be originated from the imperfect removal, or they are actually real TEOs.
Four anharmonic frequencies can pair up and sum to give exact harmonics of the orbital frequency, 
suggesting the existence of non-linear tidal processes in the eccentric binary system FX UMa.
Eight independent frequencies in the range of 20 to 32 day$^{-1}$ are typical low-order pressure modes of $\delta$ Scuti pulsators.

\end{abstract}

%% Keywords should appear after the \end{abstract} command. 
%% The AAS Journals now uses Unified Astronomy Thesaurus concepts:
%% https://astrothesaurus.org
%% You will be asked to selected these concepts during the submission process
%% but this old "keyword" functionality is maintained in case authors want
%% to include these concepts in their preprints.
. 

\section{Introduction} \label{section1}
As benchmark systems to accurately measure the masses and radii for a variety of stars with negligible model dependence, 
binary stars play a fundamental role in our understanding of stars and even the Universe \citep{Torres2010, Chen2020, Lampens2021}. 
Eclipsing binary (EB) systems hosting at least one pulsating star are much more valuable, 
since they will provide strong constraints on the input physics of asteroseismic models 
for the pulsating component and also offer the possibility to carry out mode identification 
through the technique of eclipse mapping \citep{Nather1974, Mkrtichian2018, Chen2022}. 
Meanwhile, the study of stellar oscillations will unravel the structure and dynamics of stellar interiors by means of asteroseismology \citep{Aerts2021}, 
which in turn helps us to probe the influences of tidal forces, mass transfer, and angular momentum transfer between the components \citep{Murphy2018, Guo2021, Kovalev2022}. 

Owing to the remarkable success of the space telescopes such as $\it Kepler$ \citep{Borucki2010} and $\it TESS$ \citep{Ricker2015}, 
almost all types of pulsating stars across the whole Hertzsprung-Russell Diagram, 
ranging from extremely short-period pulsating white dwarfs to massive $\beta$-Cep pulsators to long-period red giant variable stars, 
have been discovered in EB systems. 
\cite{Gaulme2019} performed a systematic search for pulsating stars in the $\it Kepler$ EB catalog\footnote{\url{http://keplerebs.villanova.edu/}} 
and identified 303 systems with stellar pulsators. 
Combining the high-precision space-based photometry and ground-based spectroscopic observations, 
researchers have made significant advances in the study of pulsating EB. 
A total of 14 double-lined spectroscopic EB with an oscillating red giant component have been found by the NASA $\it Kepler$ mission \citep{Benbakoura2021}. 
After comparing the masses and radii of the red giants given by dynamic modelling of EB systems with the results from the asteroseismic scaling relations, 
\cite{Gaulme2016} and \cite{Benbakoura2021} concluded that the asteroseismic radii and masses are systematically overpredicted by 5\% and 15\%, respectively. 
A number of detailed asteroseismic studies on individual EB systems with classical pulsators (e.g., $\delta$ Scuti and $\gamma$ Doradus) have been published 
(e.g. \citealt{Hambleton2013, Schmid2016, Guo2017, Zhang2018, Chen2021}). 
\cite{Li2020} performed an ensemble asteroseismic study of 35 $\it Kepler$ $\gamma$ Dor stars in EB systems 
and discovered that $\gamma$ Dor stars in binaries tend to show slower near-core rotation rates compared with that of single stars. 
Space missions have also contributed to the understanding of EB with compact pulsators, 
including pulsating extremely low mass white dwarfs (e.g. \citealt{Wang2020, wang2022, Kim2021}), hot subdwarfs (e.g. \citealt{Baran2021, luo2021, Dai2022}), 
and canonical white dwarfs (e.g. \citealt{Parsons2017, Garza2021}). 
Another highlight from $\it Kepler$ and $\it TESS$ is the discovery and characterisation of heartbeat stars, triggering the beginnings of tidal asteroseismology 
(e.g. \citealt{Hambleton2013, Bowman2019, Handler2020}). 
Thus far, about 180 heartbeat stars have been identified in the entire $\it Kepler$ data \citep{Kirk2016} 
and  \cite{Prsa2021} discovered 22 heartbeat binaries in the 2 minute TESS data up to Sector 26, but a handful of them have been closely studied. 
 
In this work, we use high-precision $\it TESS$ photometry and high-resolution spectroscopic measurements
 to characterize a neglected eccentric binary system FX UMa (TIC 219707463).
 In Section \ref{section2}, we describe the observations, radial velocity extraction, 
 spectroscopic orbital elements, 
 spectral disentangling, and determination of the atmospheric parameters.
The final orbital solution and physical parameters for FX UMa are given in Section \ref{section3}.
 In Section \ref{section4}, we perform a Fourier analysis of the residual light curve and discuss the pulsations. 
 Finally, we summarize our findings in Section \ref{section5}.

\section{Observations and data reductions} \label{section2} 
\subsection{TESS photometry}
\begin{figure}
\center
\includegraphics[scale=0.7]{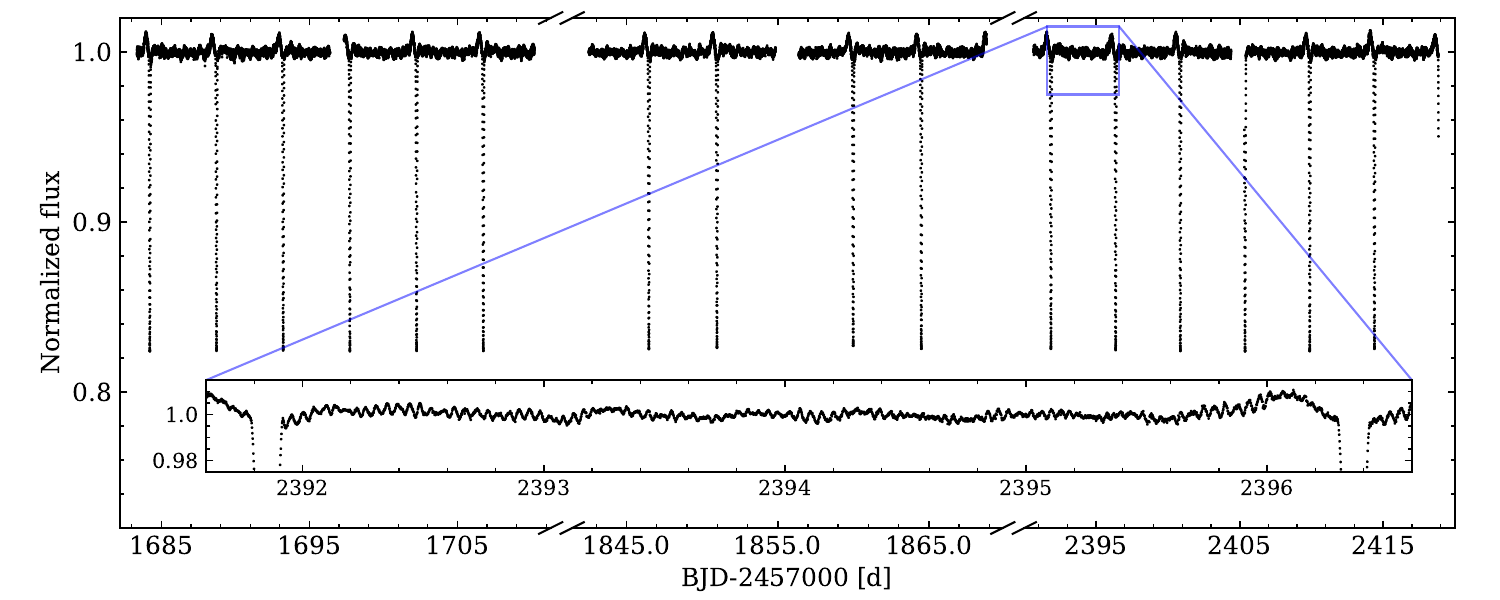}
\caption{The full 2-min $\it TESS$ light curve of FX UMa from sectors 14, 20, and 40. 
The zoom panel is a closer view of a selected region of the light curve.}
\label{fig1}
\end{figure}

Until now, FX UMa was observed by $\it TESS$ in a 2-min cadence mode during three non-contiguous sectors: 14, 20, and 40. 
The $\it TESS$ light-curve (LC) files,  produced by the $\it TESS$ Science Processing Operations Center \citep{Jenkins2016}, 
were downloaded from the Mikulski Archive for Space Telescopes (MAST\footnote{\url{https://mast.stsci.edu/portal/Mashup/Clients/Mast/Portal.html}}).
For the EB system FX UMa, we made use of the Simple Aperture Photometry data labelled SAP\underline{~~}FLUX, 
since PDCSAP\underline{~~}FLUX (Pre-Search Data Conditioning; \citealt{Smith2012}) does not detrend the LC of eclipsing binaries perfectly.
Visual examination of the raw LC reveals negligible signatures of unphysical trends.
So we didn't do anything except discard five outlying data points.
At the end, a Python package called $\it Lightkurve$\footnote{\url{https://docs.lightkurve.org/}} \citep{Lightkurve2018} was applied  to remove all nan values
and stitch together multiple sectors of $\it TESS$ observations after normalizing them.
The final $\it TESS$ light curve of FX UMa  is plotted in Figure \ref{fig1}, 
which shows a periodic brightening near the eclipse, the typical feature of heartbeat stars. 
The zoom panel presents a short segment of the LC, where the oscillations can be clearly seen.
 
\subsection{SONG high-resolution spectra}
High-resolution spectroscopic observations of FX UMa were performed by using the 1m automated Hertzsprung $\it SONG$ telescope 
at the Teide Observatory on the island of Tenerife, Spain \citep{Andersen2014, Fredslund2019}. 
The $\it SONG$ spectrograph consist of 51 spectral orders in the wavelength range of 4400 -- 6900 \AA.
The observations were carried out with slit no. 5, corresponding a spectral resolution of 77000.
We obtained a total of 34 spectra from 2019 October 28 to 2020 April 29, 
one of which was discarded in this study due to low signal-to-noise ratio (S/N). 
The 1D spectra, including the blaze function from the master (summed) flat field, were extracted with 
the $\it SONG$ spectral-reduction pipeline called $\it songwriter$ \citep{Ritter2014, Grundahl2017}.
The detailed description of the Hertzsprung $\it SONG$ telescope is well-documented in https://soda.phys.au.dk/.

\subsubsection{Radial velocity measurements}
The broadening-function technique\footnote{\url{http://www.astro.utoronto.ca/\~rucinski/SVDcookbook.html}} 
(BF; \citealt{Rucinski1992,Rucinski2002}) was used to 
determine radial velocities (RVs) from the 1D extracted spectra of FX UMa.
Compared with more familiar cross-correlation function,
 the BF method generally improves the ability to measure the Doppler shifts 
 from the complex spectra of double-lined spectroscopic binaries 
 showing substantial rotational broadening and overlapping spectral lines \citep{Rucinski2002, Bavarsad2016, Clark2019, chen101975}.
 This is essential for a short-period binary with at least one fast-rotating component star like FX UMa.
 
\begin{figure}
\center
\includegraphics[scale=0.48]{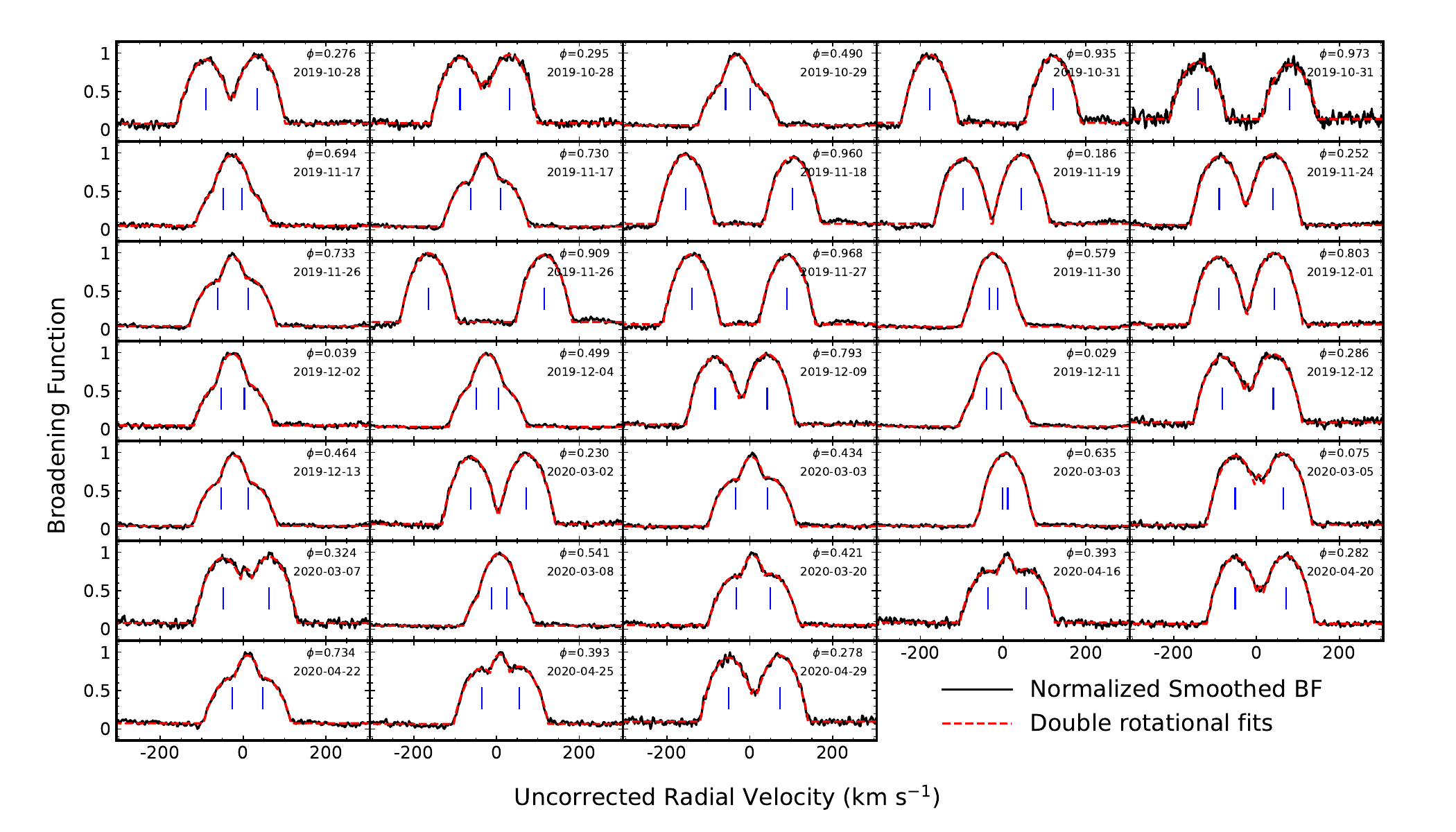}
\caption{BF plots for FX UMa. Each panel stands for one spectroscopic observation, 
for which the double-rotational profile fit to the normalized smoothed BF are plotted.
The central position of each BF, corresponding to the radial velocity of each component of FX UMa, is marked by the blue vertical line.}
\label{fig2}
\end{figure}

 We first removed the spectrograph blaze function from the 1D spectra and normalized each spectrum by its continuum order by order 
using the open-source spectroscopic tool iSpec \citep{Blanco2014, Blanco2019}.
The orders of each spectrum were then merged, 
since single spectral order covers a narrow spectral range of about 4 nm, 
which makes it hard to give strong BF peaks.
Afterwards, we computed the BFs of our target spectra by employing a modified version of the BF software suite provided publicly 
by \cite{Rawls2016}\footnote{\url{https://github.com/mrawls/BF-rvplotter}}.
A high-resolution PHOENIX synthetic spectrum \citep{Husser2013} was selected as the BF template.
The portion of spectra we computed the BFs are in the wavelength region 4900 -- 5500 \AA.
This is because this region not only staves off strong hydrogen Balmer lines, 
but also contains most of the information on the velocities. 
Following the same procedure of \cite{Rawls2016}, we adopted gauss filter to process original BF smoothly to eliminate un-correlated, small-scale noise.
The normalized smoothed BF for FX UMa is shown in Figure \ref{fig2}. 
We can see clearly from it the BF in velocity space displays two peaks,
whose positions are equal to the RVs of both components of FX UMa.
The geocentric (uncorrected) RVs, marked by the blue vertical lines in Figure \ref{fig2}, 
were obtained by rotational profile fitting to the smoothed BFs.
The barycentric velocity corrections provided by the SONG pipeline were then applied to them
to yield the final RV measurements, which are given in Table \ref{tab1}. 
The uncertainty of our measurements comes from the error in fitting a rotational profile to each BF profile 
with an open-source software package called Non-Linear Least-Squares Minimization and Curve-Fitting for Python (LMFIT)\footnote{\url{https://lmfit.github.io/lmfit-py/}}.

\subsubsection{Spectroscopic orbital elements}
\begin{figure}
\center
\includegraphics[scale=0.7]{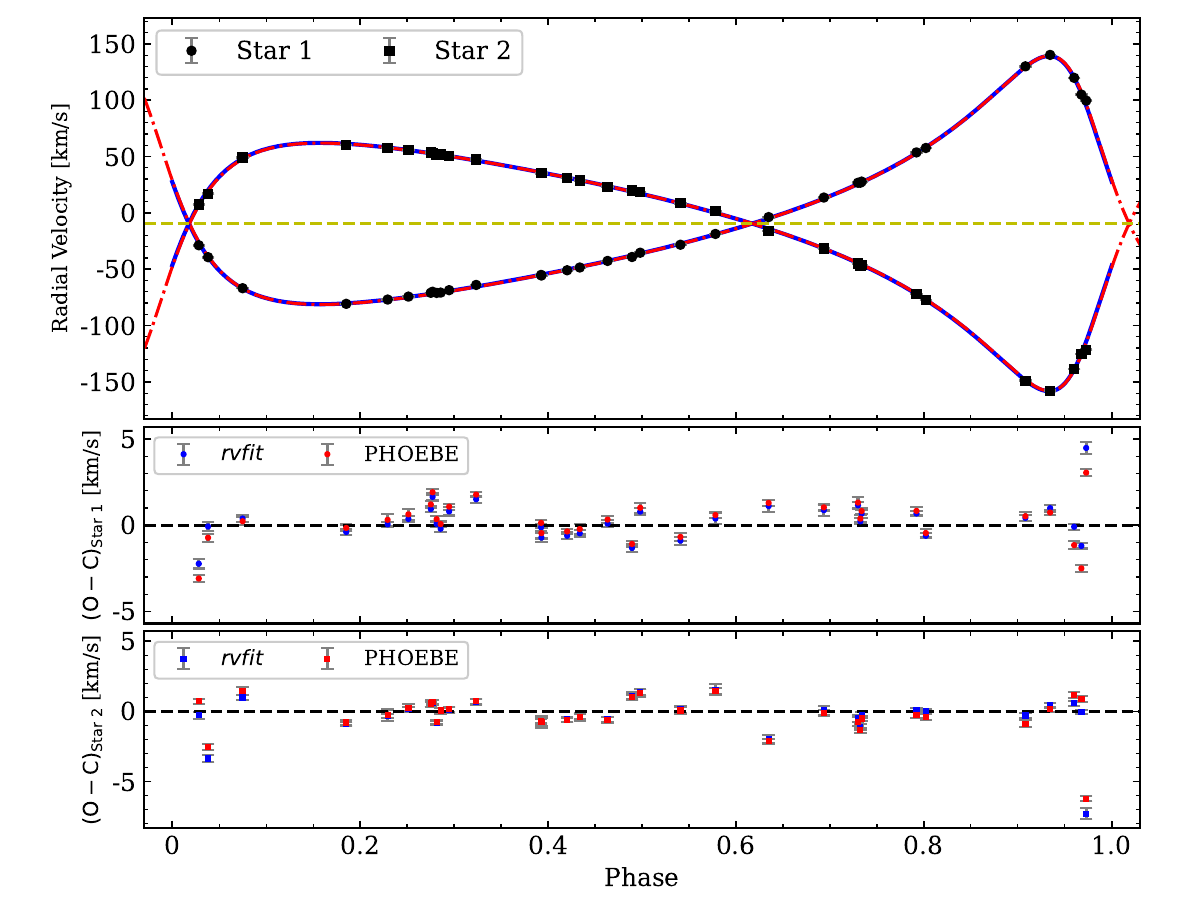}
\caption{RVs of both component stars of FX UMa as a function of phase.
The blue solid lines represent the theoretical RV curves for the both components, 
derived by using the $\it rvfit$ code \citep{Iglesias2015}. 
The red dot-dashed lines represent the best-fitting models from the PHOEBE that were constrained by both LC and RV observations,
which will be described in Section \ref{section3}.
The yellow dashed line in the top panel shows the systemic velocity of $v_{\gamma}$ = -9.475 km/s. 
The residuals from the best-fit model are presented in the bottom two panels.}
\label{fig3}
\end{figure}

In order to solve for the spectroscopic orbital parameters of FX UMa, 
we used the $\it rvfit$ code \citep{Iglesias2015} to fit our RV measurements.
Using an adaptive simulated annealing algorithm, 
the $\it rvfit$ code can automatically fit the RVs of stellar binaries and exoplanets.
It is also a user-friendly code that converges to a global solution minimum without the need to provide preliminary parameter values (see \citealt{Iglesias2015}, for details).
In the analysis of FX UMa, we fixed the orbital period to 4.50725 days, 
which was taken from the Data Validation Report Summary, provided by the $\it TESS$ Science Processing Operations Center Pipeline.
The other six orbital parameters, i.e., epoch of periastron passage ($T_{p}$), 
argument of the periastron ($\omega$), eccentricity ($e$), 
systematic velocity ($v_{\gamma}$), and semi-amplitudes of RVs for both components ($K_{1}$, $K_{2}$), were kept free during the analysis. 
The initial values for $v_{\gamma}$, $K_{1}$ and $K_{2}$ were estimated by visually examining the phased RV curves of FX UMa.
The initial values of the remaining three parameters ($T_{p}$, $\omega$, $e$) were set to 2458684.2139, 0, and 0, respectively.
To cover reasonable models, we allowed the six adjustable parameters to vary in wide ranges.
The fitted parameters and other derived quantities for the best-fit model are given in Table \ref{tab2}. 
Figure \ref{fig3} displays the theoretical RV curve fits to the observed ones and the differences between them. 
 
\begin{deluxetable}{lrrr}\label{tab1}
\tablecolumns{4}
\tablewidth{0pc}
\tabletypesize{\scriptsize}
\tablecaption{Radial velocities for FX UMa extracted from SONG spectra}
\tablehead{\colhead{BJD}    &\colhead{Phase}		&\colhead{$RV_{1}$}             &\colhead{$RV_{2}$}              \\
\colhead{(2400000+)}      		&\colhead{}		&\colhead{(km s$^{-1}$)}       &\colhead{(km s$^{-1}$)}        }
\startdata
  58784.616338  &0.276     &-70.94$\pm$0.19     	&53.47$\pm$0.18\\
  58784.704294  &0.295     &-68.58$\pm$0.24      	&50.57$\pm$0.23\\
  58785.581477  &0.490     &-39.14$\pm$0.24     	&19.96$\pm$0.27\\
  58787.587152  &0.935     &140.26$\pm$0.20     	&-157.86$\pm$0.19\\
  58787.760165  &0.973     &99.67$\pm$0.34       	&-121.47$\pm$0.40\\
  58804.530779  &0.694     &13.61$\pm$0.32    	&-31.63$\pm$0.30\\
  58804.694505  &0.730     &26.77$\pm$0.21    	&-45.09$\pm$0.22\\
  58805.731430  &0.960     &119.89$\pm$0.17   	&-138.42$\pm$0.16\\
  58806.746889  &0.186     &-80.72$\pm$0.17     	&60.49$\pm$0.15\\
  58811.552382  &0.252     &-74.31$\pm$0.15     	&55.93$\pm$0.16\\
  58813.720354  &0.733     &26.80$\pm$0.18    	&-46.66$\pm$0.19\\
  58814.512091  &0.909     &130.09$\pm$0.16   	&-148.96$\pm$0.17\\
  58814.780540  &0.968     &105.00$\pm$0.15   	&-125.25$\pm$0.14\\
  58817.532391  &0.579     &-18.67$\pm$0.31      	&1.66$\pm$0.40\\
  58818.541681  &0.803     &57.62$\pm$0.14    	&-77.23$\pm$0.16\\
  58819.605796  &0.039     &-39.30$\pm$0.27     	&16.94$\pm$0.27\\
  58821.678834  &0.499     &-35.34$\pm$0.19     	&18.55$\pm$0.21\\
  58827.511274  &0.793     &53.70$\pm$0.15    		&-71.89$\pm$0.16\\
  58828.575788  &0.029     &-28.72$\pm$0.28      	&7.28$\pm$0.26\\
  58829.736020  &0.286     &-70.76$\pm$0.21     	&51.66$\pm$0.22\\
  58830.536858  &0.464     &-42.61$\pm$0.19     	&23.21$\pm$0.20\\
  58910.612349  &0.230     &-76.92$\pm$0.15     	&57.71$\pm$0.15\\
  58911.533949  &0.434     &-48.47$\pm$0.18     	&28.70$\pm$0.19\\
  58912.439972  &0.635     &-3.72$\pm$0.34    		&-16.09$\pm$0.33\\
  58914.423942  &0.075     &-66.84$\pm$0.19     	&49.30$\pm$0.19\\
  58915.544257  &0.324     &-63.94$\pm$0.19     	&47.22$\pm$0.20\\
  58916.524161  &0.541     &-28.26$\pm$0.23      	&8.56$\pm$0.25\\
  58929.502502  &0.421     &-50.92$\pm$0.21     	&30.80$\pm$0.21\\
  58956.422493  &0.393     &-55.58$\pm$0.27     	&35.18$\pm$0.27\\
  58960.427192  &0.282     &-71.02$\pm$0.17     	&51.35$\pm$0.16\\
  58962.465837  &0.734     &27.77$\pm$0.24    	&-46.34$\pm$0.22\\
  58965.435556  &0.393     &-55.04$\pm$0.22     	&35.32$\pm$0.21\\
  58969.422138  &0.278     &-70.00$\pm$0.21     	&53.36$\pm$0.19\\
\enddata
\end{deluxetable}

\begin{deluxetable}{lr}\label{tab2}
\tablecolumns{2}
\tabletypesize{\scriptsize}
\tablecaption{Radial-velocity fitting results for the SB2 system FX UMa}
\tablehead{\colhead{Parameter}   		&\colhead{Value} }
\startdata
\multicolumn{2}{c}{Adjusted Quantities}\\
\hline
$P_{orb}$ (d)			&4.50725$^{\star}$\\
$T_p$ (BJD)		&2458688.59179 $\pm$ 0.00034\\
$e$			&0.54765 $\pm$ 0.00026\\
$\omega$ (deg)		&50.269 $\pm$ 0.049\\
$v_{\gamma}$ (km/s)	&-9.475 $\pm$ 0.024\\
$K_1$ (km/s)		&110.197 $\pm$ 0.057\\
$K_2$ (km/s)		&110.251 $\pm$ 0.056\\
\hline
\multicolumn{2}{c}{Derived Quantities}\\
\hline
$M_1\sin ^3i$ ($M_\odot$)	&1.4656 $\pm$ 0.0019\\
$M_2\sin ^3i$ ($M_\odot$)	&1.4649 $\pm$ 0.0019\\
$q = M_2/M_1$		&0.99951 $\pm$ 0.00072\\
$a_1\sin i$ ($10^6$ km)	&5.7146 $\pm$ 0.0032\\
$a_2\sin i$ ($10^6$ km)	&5.7175 $\pm$ 0.0031\\
$a  \sin i$ ($10^6$ km)	&11.4321 $\pm$ 0.0045\\
\hline
\multicolumn{2}{c}{Other Quantities}\\
\hline
$\chi^2$			&1522.83\\
$N_{obs}$ (star 1)		&33\\
$N_{obs}$ (star 2)	&33\\
Time span (days)		&184.81\\
$rms_1$ (km/s)		&1.22\\
$rms_2$ (km/s)		&1.63\\
\enddata
\tablecomments{\footnotesize $^{\star}$ Parameter fixed beforehand.}
\end{deluxetable}

\subsubsection{Atmospheric parameters from disentangled spectra}
\begin{figure}
\center
\includegraphics[scale=0.55]{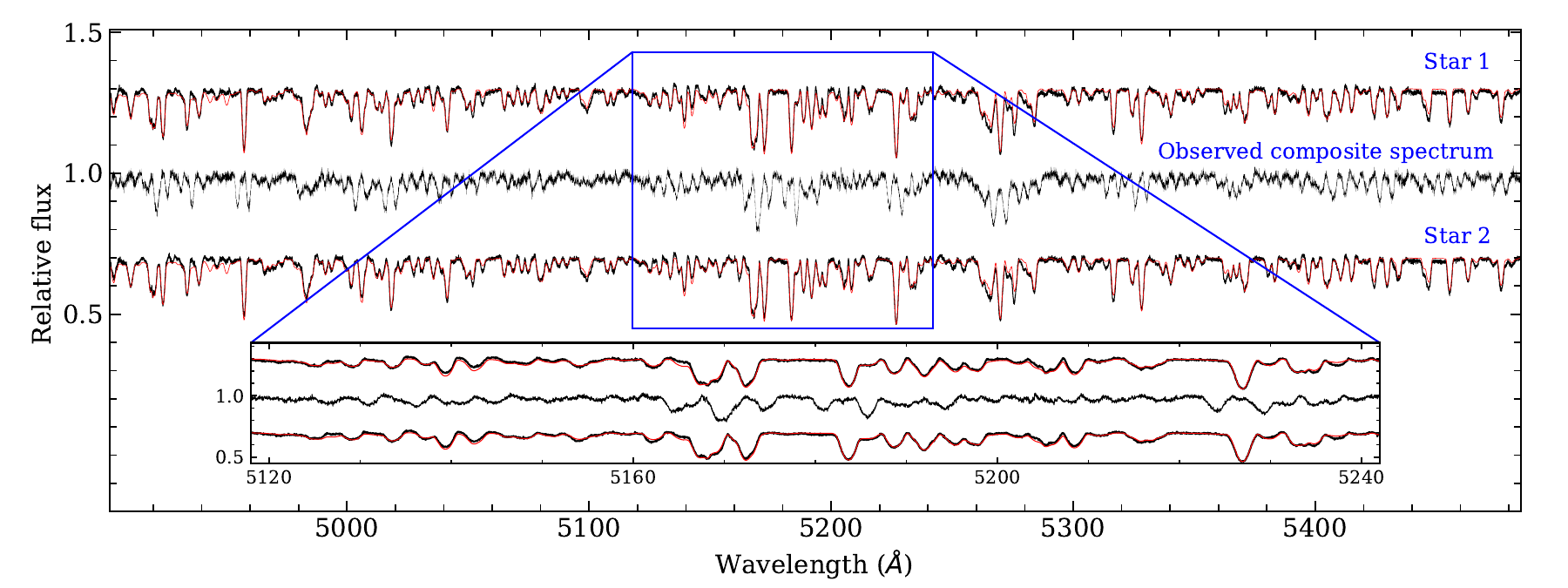}
\caption{One observed SONG spectrum (middle, BJD = 2458814.512091, $\phi$ = 0.909) and 
the disentangled spectra from FDBinary for the two components in FX UMa (upper and lower).
The zoom panel shows a small portion of these spectra to better see the details.
The besting-fitting theoretical spectra for the two stars, obtained by the iSpec code, are shown as the red solid lines.
The spectra of the stars 1 and 2 have been shifted vertically by +0.3 and -0.3, respectively, for comparison purposes.}
\label{fig4}
\end{figure}

As a double-lined spectroscopic binary system (SB2), each observed spectrum of FX UMa is a composite of individual spectra of component stars.
We attempted to reconstruct the spectra of the individual components 
using the spectral disentangling technique (for a summary of different methods see e.g., \citealt{Pavlovski2010}).
The spectral disentangling tool FDBinary\footnote{\url{http://sail.zpf.fer.hr/fdbinary/}} \citep{Ilijic2004}
was employed in this study for performing spectral decomposition. 
Without the use of template spectra, FDBinary requires six orbital parameters ($P$, $T_{p}$, $\omega$, $e$, $K_{1}$, $K_{2}$) 
to outline the shape of the Keplerian RV curve.
In the runs, we focused on the spectral interval of 4900 -- 5500 \AA\ and the orbital period ($P$) was kept at 4.50725 days.
The initial values for $T_{p}$, $\omega$, $e$, $K_{1}$, and $K_{2}$ were taken from the previous RV fitting results 
and we let them free in the analysis.
Based on the amplitude of each star's BF in Figure \ref{fig2}, the light contributions of the two component stars of FX UMa
were estimated to be 0.49 and 0.51, respectively.
One of 33 $\it SONG$ spectra that were used in RV analysis was discarded due to a relatively low S/N.
The remaining 32 $\it SONG$ spectra of FX UMa were processed together in FDBinary and 
at last we obtained the disentangled spectra of each component star, as shown in Figure \ref{fig4}.

The atmospheric parameters for both component stars were then obtained from their disentangled spectra.
All this was done using the synthetic spectra fitting technique implemented in the code iSpec \citep{Blanco2014, Blanco2019}.
iSpec integrates  a broad variety of radiative transfer codes, model atmospheres, solar abundances, and atomic line lists.
We fit the disentangled spectrum of each star separately.
In this analysis, we employed the SPECTRUM radiative code \citep{Gray1994}, 
the ATLAS9 Castelli model atmospheres \citep{Kurucz2005}, 
the Grevesse 2007 solar abundances \citep{Grevesse2007}, 
and the third version of the Vienna Atomic Line Database (VALD3; \citealt{Ryabchikova2015})
to produce synthetic spectra on the fly.
Since the surface gravity log $g$ is usually not well constrained with spectroscopy, we 
fixed log $g$ to the values derived from the stellar mass and radius (log $g_{1}$ =  log $g_{2}$ = 4.3, see the section \ref{section3}).
We iterated between the spectroscopic analysis and binary models to obtain log g.
We adopted a resolution of $R$ = 77000 appropriate to the spectral resolution of $\it SONG$ spectra.
The radial velocity for each component was set to 0, since FDBinary provides a pair of disentangled spectra with zero RV. 
In this work, we didn't analyze abundances of specific elements.
Following \cite{Blanco2014}, the limb-darkening coefficient was fixed to 0.6.
The adjustable parameters consist of effective temperature $T_{eff}$, metallicity $[M/H]$,
alpha enhancement $[\alpha/Fe]$, microturbulence velocity $v_{mic}$,
and projected rotational velocity $v_{rot}$sin$i$.
The application can automatically calculate the macroturbulence velocity 
based on an empirical relation established by Gaia-ESO Survey.
The resulting atmospheric parameters for both stars in FX UMa are presented in Table \ref{tab3}. 
Figure \ref{fig4} displays the observed composite, disentangled, and fitted synthetic spectra 
of both components of FX UMa. 

\section{Binary modelling} \label{section3} 
As seen in Figure \ref{fig1}, the LCs of FX UMa show a periodic, broad brightening near the eclipse,
which look distinctly different from intrinsic star variability and instead are characteristic of heartbeat stars. 
Apart from the eclipse-like light changes and the ``heartbeat-like" profile, the LCs outside of eclipses present 
multi-periodic light variations with characteristic of hybrid $\delta$ Sct - $\gamma$ Dor oscillations.
To obtain the physical parameters of FX UMa and probe its pulsational properties in detail, 
we carried out a simultaneous fit to our double-lined RVs and $\it TESS$ LCs
using the PHysics Of Eclipsing BinariEs (PHOEBE, \citealt{Andrej2018, Conroy2020}) 
code\footnote{\url{http://phoebe-project.org/}} in the detached mode. 

\subsection{PHOEBE setup}
The initial values for the mass ratio ($q$ = $M_{2} / M_{1}$), orbital eccentricity ($e$), argument of periastron ($\omega$), 
projected orbital semi-major axis ($a_{orb}$ sin$i$), and systemic velocity ($v_{\gamma}$), were taken from Table \ref{tab2}.
The input effective temperatures for both components ($T_{\rm eff, 1}$,  $T_{\rm eff, 2}$) were taken from our spectral analysis. 
The orbital ephemeris, i.e. reference time of superior conjunction ($t_{0}$) and orbital period ($P_{orb}$), 
were obtained by the TESS light curves of FX UMa.  
The initial value for the synchronicity parameter was computed by using the formalism of $F_{1, 2}$ = $\sqrt{(1~+~e)/(1~ - ~e)^3}$ \citep{Andrej2018}, 
where $e$ is the orbital eccentricity from the above RV fit.
The limb-darkening coefficients for each star were automatically interpolated from the PHOEBE's built-in lookup table, 
with the atmosphere models set as PHOENIX. 
The free parameters in our model were: 
$t_{0}$, $P_{orb}$, $q$, $e$, $\omega$, $a_{orb}$ sin$i$, $v_{\gamma}$, $T_{\rm eff, 1}$,  the temperature ratio ($T_{\rm eff, 2}$ / $T_{\rm eff, 1}$), 
the orbital inclination ($i$), the sum and ratio of fractional radii (($R_{\rm equiv, 1}$ + $R_{\rm equiv, 2}$) / $a_{orb}$, $R_{\rm equiv, 2}$ / $R_{\rm equiv, 1}$, 
where $a_{orb}$ is the semi-major axis of the orbit), 
the gravity-brightening coefficient ($g_{1,2}$), the bolometric albedo ($A_{1,2}$), the third light ($l_{3}$), and the passband luminosity of star 1 ($L_{pb}$).  
We utilized the emcee sampler \citep{Foreman2019, Foreman2013} built into PHOEBE to 
explore the parameter spaces, find the optimal solution, and determine the uncertainties.
We used 160 walkers with chain lengths of 5000 each, resulting in a total of 800,000 model computations.
Convergence was checked  both by visual examination of the chains and by inspecting the autocorrelation times for all the fitted parameters.
 
\subsection{Uncertainties and stellar parameters} 
The preliminary solutions resulted in unrealistic errors for the two component stars.
For example, the mass uncertainty is approximately equal to 0.0003$M_{\sun}$.
We suspect that the observational uncertainties in both the radial velocities and light curve may have been underestimated. 
These values feed into the uncertainties on the parameters and are likely the cause, at least in part, of the posteriors showing underestimated uncertainties.
It is evident from Figure \ref{fig3} that there have been significant underestimations in the measurement errors of our radial velocities.
  
We refined the uncertainty estimates for the radial velocities and light curve by using the residuals obtained from removing the PHOEBE model fit from the original observational data.
The standard deviations of the RV residuals for the two component stars were calculated to be $\sigma_{rv,1}$ $\simeq$ 1.170 km/s and $\sigma_{rv,2}$ $\simeq$ 1.379 km/s, respectively.
The median values of the RV observational uncertainties were found to be 0.202 km/s and 0.204 km/s for the two components, respectively.
The standard deviation of the residuals for the normalized fluxes was calculated to be $\sigma_{lc}$ $\simeq$ 0.001546 and the median value of the LC observational errors was found to be 0.000204.
In order to ensure that the PHOEBE model can traverse the parameter space thoroughly, 
we used the 3$\sigma$ values of the RV and LC residuals as the typical errors for each dataset and then divided them by the corresponding median values of the original observational errors.    
Therefore, the errors for the radial velocities of the component stars 1 and 2 were increased to 17.4 and 20.3 times of their original values, respectively.
The uncertainties of the light curve were amplified to 22.8 times of their original values.  
In the end, we performed a new EMCEE analysis on the light curve and radial velocities with adjusted uncertainties.

\begin{figure}
\center
\includegraphics[scale=0.65]{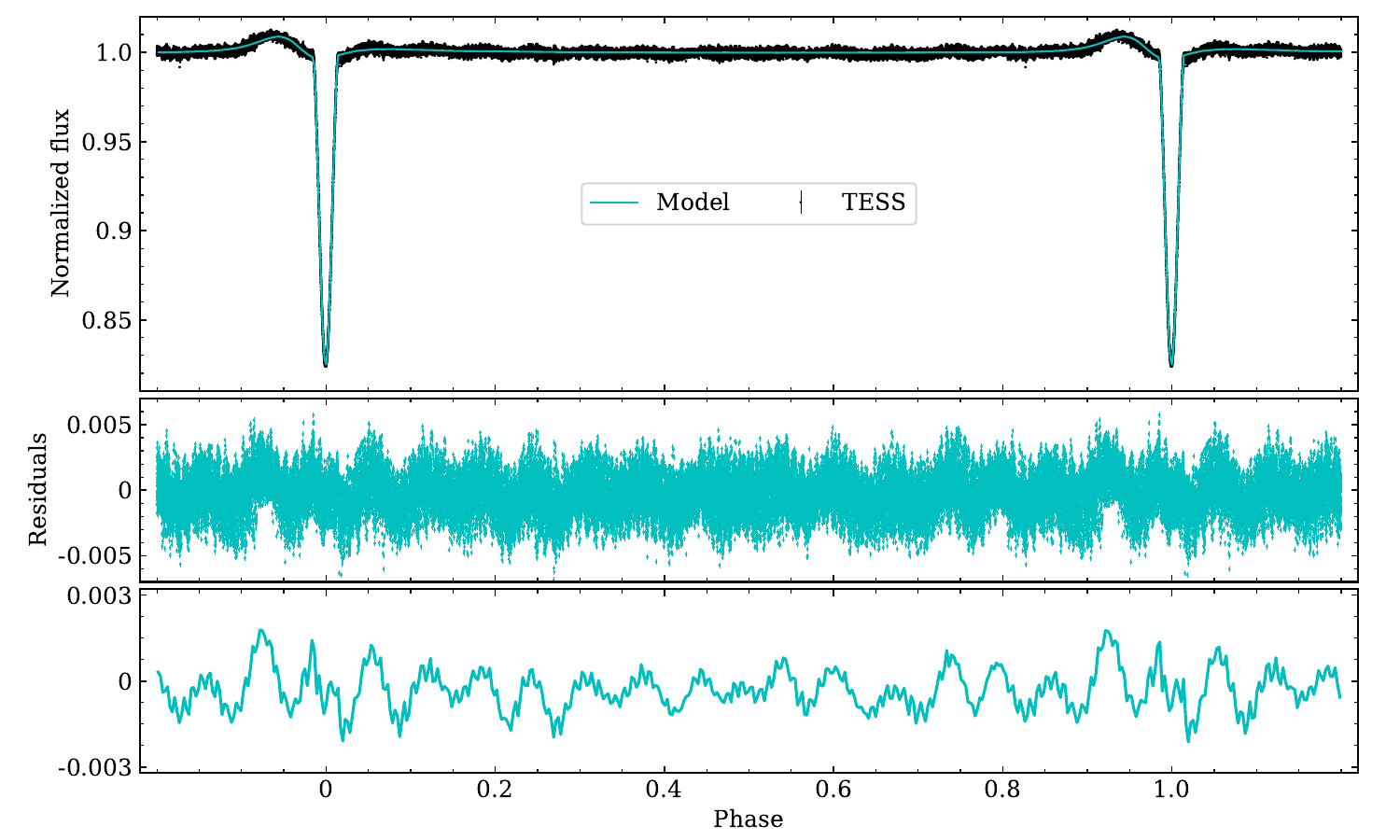}
\caption{\bf The top panel shows the phased TESS light curve of FX UMa with the best-fitting model superimposed.
The unbinned and binned light residuals are presented in the middle and bottom panels, respectively.}
\label{fig5}
\end{figure}

The median value of each parameter's posterior distribution is reported in Table \ref{tab3}, 
in which the upper and lower uncertainties are obtained at the 16th and 84th percentiles, respectively. 
The synthetic RV curves for the two stars, produced by the final PHOEBE binary model, are overplotted in Figure \ref{fig3}.
The best fit light curve is shown as solid line in the top panel of Figure \ref{fig5}. 
The non-binned and binned light residuals are displayed in the middle and bottom panels, respectively.   
Figure \ref{fig5} illustrates that there are few systematic trends in the LC residuals. 
Both the light curve and radial velocity curves are fairly well matched.
In Appendix, we show the parameter posterior distributions and their interdependencies for the PHOEBE model.
 
\begin{deluxetable}{lrr}\label{tab3}
\tablecolumns{3}
\tablewidth{0pc}
\tablecaption{Atmospheric parameters and binary model parameters for FX UMa}
\tablehead{\multicolumn{3}{c}{iSpec analysis}}
\startdata
Parameters    &Star 1    		    		&Star 2\\
\hline
$T_{eff}$ (K)		         	&7392$\pm$44				&7390$\pm$44	\\
log $g$ (dex)				&4.3$^{\star}$				&4.3$^{\star}$		\\
$[M/H]$ (dex)				&-0.41$\pm$0.02			&-0.43$\pm$0.02	\\
$[\alpha/Fe]$ (dex)			&0.23$\pm$0.03			&0.23$\pm$0.03	\\
$v_{mic}$ (km s$^{-1}$)		&4.34$\pm$0.12			&4.29$\pm$0.12	\\
$v_{rot}$sin$i$ (km s$^{-1}$)	&65.4$\pm$0.8				&66.8$\pm$0.8	\\		
\hline
\multicolumn{3}{c}{PHOEBE analysis$^{\dagger}$}\\
\hline
Parameters									&\multicolumn{2}{r}{Value}\\
\hline
$t_{0}$			  	  			 	 		&\multicolumn{2}{r}{1684.21622$_{-5e-5}^{+5e-5}$}	\\
$P_{orb}$ (days)                				 		&\multicolumn{2}{r}{4.5072522$_{-5e-7}^{+5e-7}$}	\\
$i$ (deg)                         					 		&\multicolumn{2}{r}{78.63$_{-0.06}^{+0.07}$}\\
$e$									  		&\multicolumn{2}{r}{0.547$_{-0.002}^{+0.002}$}\\
$\omega$ (deg)						  		&\multicolumn{2}{r}{50.3$_{-0.3}^{+0.4}$}	\\
$v_{\gamma}$ (km s$^{-1}$)			           		&\multicolumn{2}{r}{-9.5$_{-0.4}^{+0.3}$}\\
$q=M_{2}/M_{1}$                 				  		 &\multicolumn{2}{r}{1.004$_{-0.007}^{+0.008}$}\\ 
$a_{orb}$ sin$i$ ($R_{\sun}$) 				   		&\multicolumn{2}{r}{16.44$_{-0.08}^{+0.09}$}\\
$T_{\rm eff,1}$ (K)						   		&\multicolumn{2}{r}{7413$_{-32}^{+49}$}\\
$T_{\rm eff,2}/T_{\rm eff,1}$				   		&\multicolumn{2}{r}{0.985$_{-0.010}^{+0.009}$}\\
($R_{\rm equiv, 1}$ + $R_{\rm equiv, 2}$) / $a_{orb}$	&\multicolumn{2}{r}{0.1796$_{-0.0006}^{+0.0006}$}\\
$R_{\rm equiv, 2}$ / $R_{\rm equiv, 1}$				&\multicolumn{2}{r}{0.97$_{-0.02}^{+0.04}$}\\
$L_{pb}$ 										&\multicolumn{2}{r}{6.6$_{-0.3}^{+0.2}$}\\
$l_{3}$										&\multicolumn{2}{r}{0.008$_{-0.003}^{+0.003}$}\\
$F_{1}$										&\multicolumn{2}{r}{4.8$_{-0.3}^{+0.4}$}\\
$F_{2}$										&\multicolumn{2}{r}{4.7$_{-0.5}^{+1.0}$}\\
$A_{1}$										&\multicolumn{2}{r}{0.94$_{-0.04}^{+0.04}$}\\
$A_{2}$										&\multicolumn{2}{r}{0.95$_{-0.04}^{+0.04}$}\\
$g_{1}$										&\multicolumn{2}{r}{0.93$_{-0.04}^{+0.05}$}\\
$g_{2}$										&\multicolumn{2}{r}{0.90$_{-0.03}^{+0.03}$}\\
\hline
\multicolumn{3}{c}{Stellar parameters derived by the PHOEBE code}\\
\hline
Parameters    &Star 1    		    		&Star 2\\
\hline
$M$ ($M_{\sun}$) 		&1.55$_{-0.02}^{+0.02}$ 				&1.56$_{-0.03}^{+0.03}$\\
$R$ ($R_{\sun}$)              &1.53$_{-0.04}^{+0.03}$               		&1.49$_{-0.02}^{+0.03}$\\
$log$ $g$ (dex)			&4.26$_{-0.01}^{+0.02}$				&4.29$_{-0.02}^{+0.01}$\\
\enddata
\end{deluxetable}

\section{Pulsation characteristics} \label{section4}
\begin{figure}
\center
\includegraphics[scale=0.8]{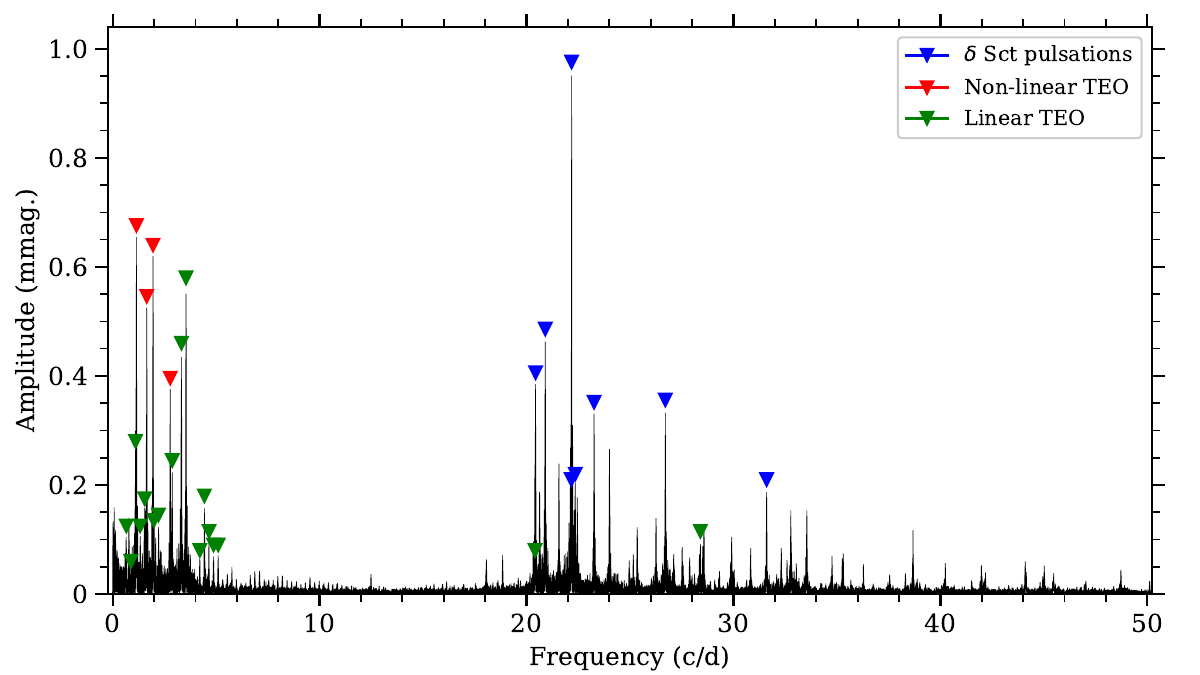}
\caption{The amplitude spectrum of the residual LC of FX UMa after subtracting the binarity-induced light variations.
The red, green, and blue upside-down triangles mark the frequencies of the detected non-linear TEOs, linear TEOs, and 
$\delta$ Sct pulsations, respectively.
The remaining significant peaks are the combination frequencies, as summarized in Table \ref{tb5}.}
\label{fig6}
\end{figure}

The residual LC of FX UMa was obtained by removing the modeled EB light curve from the original TESS observations.
We utilized the Period04 software \citep{Lenz2005} to search for frequencies of pulsation in the residual LC of FX UMa.
This was done through an iterative pre-whitening process. 
We stopped the frequency search when in the Fourier amplitude spectrum none of peaks satisfied S/N ratio $\geq$ 4 rule.
The search was restricted in the frequency range from 0 to 80 day$^{-1}$.
We have also checked for frequencies up to the Nyquist frequency of two-minute cadence TESS data ($\simeq$ 359 day$^{-1}$), 
but did not find any peaks beyond 80 day$^{-1}$.
Finally, we detected a total of 103 frequencies with S/N $\geq$ 4.
With the frequency resolution of 0.002 day$^{-1}$, we performed a search for potential 
orbital harmonics ($f_{i}$ = $Nf_{orb}$, $f_{orb}$ = 0.22186467$\pm$0.00000001 day$^{-1}$) and combination frequencies 
using our own codes and Period04, respectively. 
We identified 12 independent frequencies and 17 multiples of the orbital frequency, both of which are listed in Table \ref{tb4}.
The remaining 74 frequencies presented in Table \ref{tb5} are found as probable combination frequencies.
Figure \ref{fig6} shows the Fourier amplitude spectrum for the residual LC of FX UMa. 

\subsection{Tidally Excited Modes}
More than 20\% heartbeat stars have been observed to show tidally excited oscillations (TEOs, \citealt{Kurtz2022}), driven by dynamical tides.
Most of the observed TEOs occur at harmonics of the orbital frequency ($f_{i}$ = $Nf_{orb}$), 
which are likely triggered by the linear dynamical tide.
That is to say, the detected $Nf_{orb}$ peaks are thought to the signature of tidally excited modes.
At the same time we recognize that the imperfect removal of the binarity-induced light variations
can result in alias peaks of the form $Nf_{orb}$ with low amplitudes. 
As seen in Table \ref{tb4}, there are a total of 17 orbital frequency harmonics 
in the range of 0.6 to 28.4 day$^{-1}$ ($N$ = 3 to 128).
Thereinto, the most prominent peak is the $f_{4}$ = 3.549841$\pm$0.000003 day$^{-1}$ $\simeq$ 16$f_{orb}$
with an amplitude of 0.562 mmag and a S/N of 80.3, which can not be attributed to imperfect light curve modeling. 
Following \cite{Guo2019}, we consider the 10 $Nf_{orb}$ peaks with S/N $>$ 10 as high probability TEOs.
However, we cannot rule out the possibility that other $Nf_{orb}$ peaks (4 $\leq$ S/N $\leq$ 10) are actually real TEOs.        

In the lower frequency $g$-mode regime, the four significant frequencies of $f_{2}$, $f_{3}$, $f_{5}$, and $f_{9}$
are not multiples of the $f_{orb}$ and may be self-excited $\gamma$ Dor-type $g$ modes.
After checking whether there is any connection between these anharmonic frequencies and the orbital frequency,
we found that they can pair up and sum to give exact harmonics of the orbital frequency: 
$f_{5}$ + $f_{9}$ $\simeq$ $f_{20}$ $\simeq$ 20$f_{orb}$  and $f_{2}$ + $f_{3}$ $\simeq$ 14$f_{orb}$.
This implies the existence of non-linear tidal processes in the eccentric binary system FX UMa.
When the amplitude of a linear TEO mode exceeds the parametric instability threshold,
then it may experience non-linear resonance mode coupling and decay into (or more) daughter modes \citep{Weinberg2012, Yu2020}.
Observationally, the sum of the daughters's frequencies is equal to the frequency of the parent mode.
Non-linear tidal oscillations have been found in some eccentric binary systems, 
such as KOI-54 \citep{Burkart2012,Guo2022}, KIC 4544587 \citep{Hambleton2013},
KIC 3858884 \citep{Manzoori2020}, and KIC 3230227 \citep{Guo2020}.
Therefore, we argue that $f_{2}$, $f_{3}$, $f_{5}$, and $f_{9}$
can be considered as the non-linearly excited daughter modes of different parent modes.
The parent mode of $f_{5}$ and $f_{9}$ is resonantly driven by a linear dynamical tide at $f_{20}$ $\simeq$ 20$f_{orb}$.
Interestingly, we did not detect the parent mode of $f_{2}$ and $f_{3}$, which was supposed to be at 14 times the orbital frequency.
The same situation exists for the eccentric binary system KIC 4544587 \citep{Hambleton2013}.
The authors concluded that the two daughter modes probably come from non-linear driving by the equilibrium tide.
Following \citep{Hambleton2013}, the parent mode of $f_{2}$ and $f_{3}$ is the component of the equilibrium tide 
that pulsates at an orbital harmonic of 14$f_{orb}$.

\subsection{$\delta$ Scuti-type $p$ Modes}
In the high-frequency region, we obtained 8 independent frequencies ranging from 20 to 32 day$^{-1}$.
These frequencies are typical $\delta$ Sct-type pressure modes.
The two component stars of FX UMa are almost identical 
and their physical properties agree well with those of a typical $\delta$ Scuti pulsator.
So it is very hard to ascertain which star the observed $\delta$ Sct-type pulsations originated from. 
We got the pulsation constants ($Q$) of all these modes using the physical parameters of the star 1 and the equation of $Q=P_{pul}(\bar{\rho}_{1}/\bar{\rho}_{\sun})^{1/2}$, 
where $P_{pul}$ is the pulsation period and $\bar{\rho}_{1}$ is the mean density, $\bar{\rho}_{1}$ = $M_{1}$/(4$\pi$$R_{1}$$^3$/3).
The $Q$ values are in the range of 0.021 to 0.033 days, suggesting low-order $p$ mode oscillations of $\delta$ Sct stars \citep{Breger2000}.

%\startlongtable
\begin{deluxetable*}{lrrrrcc}\label{tb4}
\tablecolumns{7}
\tablewidth{0pc}
\tablecaption{Independent oscillation frequencies and orbital harmonic frequencies for FX UMa}
\tablehead{\colhead{ID}   &\colhead{Frequency}   &\colhead{Amplitude}  &\colhead{Phase}   &\colhead{S/N}  &\colhead{Remark}  &\colhead{Comment}\\
\colhead{}    &\colhead{(day$^{-1}$)}     &\colhead{(mmag)}   &\colhead{(rad/2$\pi$)}   &\colhead{}    &\colhead{} &\colhead{}}
\startdata
$f_{1}$      &22.176698$\pm$0.000045     	&0.913$\pm$0.039    	&0.658$\pm$0.006    	&89.1        &\nodata		&$\delta$ Sct  \\     
$f_{2}$	 &1.152015$\pm$0.000004	&0.629$\pm$0.007		&0.117$\pm$0.001		&55.1	&\nodata		&Non-linear TEO\\
$f_{3}$	 &1.954066$\pm$0.000003	&0.625$\pm$0.004		&0.508$\pm$0.001		&59.4	&\nodata		&Non-linear TEO\\ 
$f_{4}$	 &3.549841$\pm$0.000003	&0.562$\pm$0.003		&0.722$\pm$0.001		&80.3	&16$f_{orb}$	&Linear TEO\\
$f_{5}$	 &1.651120$\pm$0.000005	&0.528$\pm$0.007		&0.713$\pm$0.001		&47.9	&\nodata		&Non-linear TEO\\
$f_{6}$	 &20.903592$\pm$0.000011	&0.458$\pm$0.014		&0.775$\pm$0.003		&52.2	&\nodata		&$\delta$ Sct	\\
$f_{7}$	 &3.327977$\pm$0.000005	&0.454$\pm$0.003		&0.368$\pm$0.001		&60.2	&15$f_{orb}$	&Linear TEO\\
$f_{8}$	 &20.439244$\pm$0.000036	&0.378$\pm$0.017		&0.511$\pm$0.008		&47.1	&\nodata		&$\delta$ Sct	\\
$f_{9}$	 &2.786152$\pm$0.012667	&0.394$\pm$0.103		&0.255$\pm$0.264		&45.4	&\nodata		&Non-linear TEO	\\
$f_{10}$	 &26.708523$\pm$0.000022	&0.324$\pm$0.011		&0.418$\pm$0.006		&38.3	&\nodata		&$\delta$ Sct	\\
$f_{11}$	 &23.263153$\pm$0.000005	&0.327$\pm$0.004		&0.764$\pm$0.001		&33.6	&\nodata		&$\delta$ Sct	\\
$f_{14}$ 	 &2.884247$\pm$0.000012	&0.228$\pm$0.006		&0.681$\pm$0.003		&27.0	&13$f_{orb}$	&Linear TEO\\
$f_{16}$	 &1.109313$\pm$0.002787	&0.212$\pm$0.027		&0.219$\pm$0.175		&18.5	&5$f_{orb}$	&Linear TEO\\
$f_{17}$	 &22.161595$\pm$0.008500	&0.194$\pm$0.044		&0.978$\pm$0.169		&19.0	&\nodata		&$\delta$ Sct	\\
$f_{18}$	 &22.352218$\pm$0.000016	&0.194$\pm$0.007		&0.789$\pm$0.005		&18.5	&\nodata		&$\delta$ Sct	\\  
$f_{19}$	 &31.599018$\pm$0.019344	&0.190$\pm$0.039		&0.144$\pm$0.107		&28.2	&\nodata		&$\delta$ Sct	\\
$f_{20}$	 &4.437298$\pm$0.000011	&0.153$\pm$0.004		&0.042$\pm$0.003		&24.7	&20$f_{orb}$ $\simeq$ $f_{5}$ + $f_{9}$ &Linear TEO \\
$f_{24}$	 &1.553085$\pm$0.002818	&0.141$\pm$0.026		&0.150$\pm$0.234		&12.7	&7$f_{orb}$	&Linear TEO\\
$f_{32}$	 &2.218628$\pm$0.000017	&0.121$\pm$0.004		&0.258$\pm$0.004		&11.6	&10$f_{orb}$	&Linear TEO\\  
$f_{36}$	 &0.665672$\pm$0.000021	&0.098$\pm$0.004		&0.592$\pm$0.007		&8.5		&$3f_{orb}$	&Probable TEO\\
$f_{40}$	 &1.331203$\pm$0.016790	&0.097$\pm$0.022		&0.655$\pm$0.173		&8.5		&6$f_{orb}$	&Probable TEO\\
$f_{41}$	 &1.996715$\pm$0.000021	&0.093$\pm$0.004		&0.048$\pm$0.006		&8.9		&$9f_{orb}$	&Probable TEO\\
$f_{42}$	 &28.399938$\pm$0.004343	&0.087$\pm$0.013		&0.816$\pm$0.116		&9.5		&128$f_{orb}$ $\simeq$ $f_{10}$ + $f_{11}$ - $f_{13}$	&Probable TEO\\
$f_{43}$	 &4.659156$\pm$0.000016	&0.093$\pm$0.003		&0.928$\pm$0.006		&14.2	&$21f_{orb}$	&Linear TEO\\
$f_{64}$	 &4.881017$\pm$0.000022	&0.067$\pm$0.003		&0.249$\pm$0.008		&10.6	&22$f_{orb}$	&Linear TEO\\
$f_{65}$	 &5.102835$\pm$0.000025	&0.067$\pm$0.004		&0.676$\pm$0.006		&10.7	&23$f_{orb}$	&Linear TEO\\
$f_{74}$	 &4.215388$\pm$0.000023	&0.060$\pm$0.004		&0.347$\pm$0.008		&9.9		&$19f_{orb}$	&Probable TEO\\
$f_{78}$	 &0.887649$\pm$0.013698	&0.061$\pm$0.014		&0.798$\pm$0.175		&5.2		&$4f_{orb}$	&Probable TEO\\
$f_{97}$	 &20.411762$\pm$0.077050	&0.061$\pm$0.059		&0.733$\pm$0.280		&7.7		&92$f_{orb}$ $\simeq$ $f_{11}$ + $f_{37}$ - $f_{14}$	&Probable TEO\\
\enddata
\end{deluxetable*}

\section{Summary}\label{section5}
 We report the first result of the study of pulsating eclipsing binaries 
 combining high-precision $\it TESS$ photometry and high-resolution $\it SONG$ spectroscopic observations.
 In this work, we have carried out a detailed analysis of a bright double-lined spectroscopic binary FX UMa.
The major results can be summarized as follows:
\begin{enumerate}
\item FX UMa was observed by $\it TESS$ in 2 minutes cadence mode during three non-contiguous sectors: 14, 20, and 40.
The $\it TESS$ light curve shows a periodic, broad brightening near the eclipse, the typical feature of heartbeat stars. 
In addition to the eclipse-like light changes and the “heartbeat-like” profile, the LC in the outside eclipses clearly display multi-periodic light variations.

\item We obtained a total of 33 $\it SONG$ high-resolution spectra from 2019 October 28 to 2020 April 29.
$\it SONG$ spectra confirm that FX UMa is a double-lined spectroscopic binary system.
The radial velocities of the two component stars were extracted by using the broadening-function technique.
A joint modelling of TESS light curve and SONG radial-velocity measurements yields a mass ratio of $q$ = 1.004$_{-0.007}^{+0.008}$, 
and a high eccentricity of $e$ = 0.547$_{-0.002}^{+0.002}$, for this binary system.

\item We reconstructed the individual spectra of each component star from the observed composite spectra 
with the spectral disentangling tool FDBinary.
The atmospheric parameters for both component stars were then obtained through fitting their disentangled spectra.
The two components of FX UMa have almost exactly the same atmospheric parameters,
 with $T_{eff}$ = 7391.7$\pm$43.7 K, $[M/H]$ = -0.41$\pm$0.02 dex, $v_{rot}$sin$i$ = 65.40$\pm$0.80 km s$^{-1}$ for star 1
 and $T_{eff}$ = 7389.8$\pm$43.9 K, $[M/H]$ = -0.43$\pm$0.02 dex, $v_{rot}$sin$i$ = 66.79$\pm$0.83 km s$^{-1}$ for star 2.
 
 \item We performed a simultaneous fit to our double-lined RVs and $\it TESS$ light curves with the PHOEBE code. 
 The fitting results indicate that FX UMa is a detached, eccentric binary system with 
 an inclination of about 78.63 degrees. 
The derived physical parameters for this binary are as follows:
 $M_{1}$ = 1.55$_{-0.02}^{+0.02}$, 
$R_{1}$ =  1.53$_{-0.04}^{+0.03}$,
and $M_{2}$ = 1.56$_{-0.03}^{+0.03}$, 
$R_{2}$ = 1.49$_{-0.02}^{+0.03}$.
 This means FX UMa is an eclipsing binary with twin component stars.
 Such systems have been recently found in several heartbeat stars, such as KOI-54 \citep{Burkart2012} and KIC 4142768 \citep{Guo2019}.
 
 \item We utilized the Period04 software to extract significant frequencies from the residual LC of FX UMa, 
 obtained by removing the modeled EB light curve from the original TESS observations.
 We detected a total of 103 frequencies with S/N $\geq$ 4, including 12 independent frequencies, 
 17 multiples of the orbital frequency, and 74 combination frequencies.
 The eight independent frequencies in the range of 20 to 32 day$^{-1}$
 are typical low-order pressure modes of $\delta$ Scuti pulsators.
 At present it is hard to find out which star the observed $\delta$ Sct-type pulsations originated from, since the two components of FX UMa are almost identical.
 Most of the observed TEOs oscillate at harmonics of the orbital frequency ($f_{i}$ = $Nf_{orb}$).
 Consequently, $Nf_{orb}$ peaks have been viewed as the signature of tidally excited modes triggered by the linear dynamical tide.
The ten $Nf_{orb}$ peaks with S/N $>$ 10 have very high amplitudes and are considered as high-probability TEOs. 
The remaining $Nf_{orb}$ peaks (4 $\leq$ S/N $\leq$ 10)  may be originated from the imperfect removal, 
or they are actually real TEOs. We found that the four anharmonic frequencies ($f_{2}$, $f_{3}$, $f_{5}$, and $f_{9}$)
can pair up and sum to give exact harmonics of the $f_{orb}$: 
$f_{5}$ + $f_{9}$ $\simeq$ $f_{20}$ $\simeq$ 20$f_{orb}$  and $f_{2}$ + $f_{3}$ $\simeq$ 14$f_{orb}$. 
They are probably attributed to the non-linearly excited daughter modes of different parent modes
that are resonantly driven by the linear dynamical tide.
\end{enumerate} 

Heartbeat stars with TEOs provide unique opportunities to test theories of stellar tides and their interaction with pulsation, 
and with angular momentum \citep{Kurtz2022}.
As summarized in Table 1 of \cite{Guo2021}, 22 heartbeat binaries have been observed to show tidally excited oscillations, 
but only a handful of them have been studied in detail. 
The discovery of linear and non-linear tidal oscillations in the SB2 system FX UMa 
presents us with a new opportunity. 
During our analysis, additional $\it TESS$ observations for this object were released for Sector 47.
However, the two-minute cadence data of FX UMa in Sector 47 were not available at the time of publication.
Based on the Web $\it TESS$ Viewing Tool \footnote{\url{https://heasarc.gsfc.nasa.gov/cgi-bin/tess/webtess/wtv.py}}, 
we observe further that FX UMa will be observed by $\it TESS$ during Sector 60.
Long time series of $\it TESS$ photometry help to resolve individual pulsations. 
We expect more TEOs will be reported in the future.

\begin{acknowledgments}
This work is supported by the National Natural Science Foundation of China (Grant No. 12003022, 12173028), 
the Sichuan Science and Technology Program (Grant No. 2020YFSY0034), 
the National Key Basic R\&D Program of China (Grant No. 2021YFA1600401),
the Sichuan Youth Science and Technology Innovation Research Team (Grant No. 21CXTD0038),
and the Innovation Research Team funds of China West Normal University (Grant No. KCXTD2022-6). 
R.A.B acknowledges the support from the Major Science and Technology Project of Qinghai Province (Grant No. 2019-ZJ-A10).
This research is based on observations made at the Hertzsprung SONG telescope 
operated at the Spanish Observatorio del Teide on the island of Tenerife 
by the Aarhus and Copenhagen Universities and by the Instituto de Astrofísica de Canarias.
Funding for the Stellar Astrophysics Centre is provided by The Danish National Research Foundation (Grant agreement no.: DNRF106)

\end{acknowledgments}

\software{$\it Lightkurve$ \citep{Lightkurve2018},
iSpec \citep{Blanco2014, Blanco2019},
BF-rvplotter (\url{https://github.com/mrawls/BF-rvplotter}),
LMFIT \citep{Newville2021},
$\it rvfit$ \citep{Iglesias2015},
FDBinary \citep{Ilijic2004},
PHOEBE, \citep{Andrej2018, Conroy2020},
emcee \citep{Foreman2019, Foreman2013}, 
Period04 \citep{Lenz2005}.
}

\bibliography{aasbibtex}{}
\bibliographystyle{aasjournal}

\begin{figure}
\center
\includegraphics[scale=0.25]{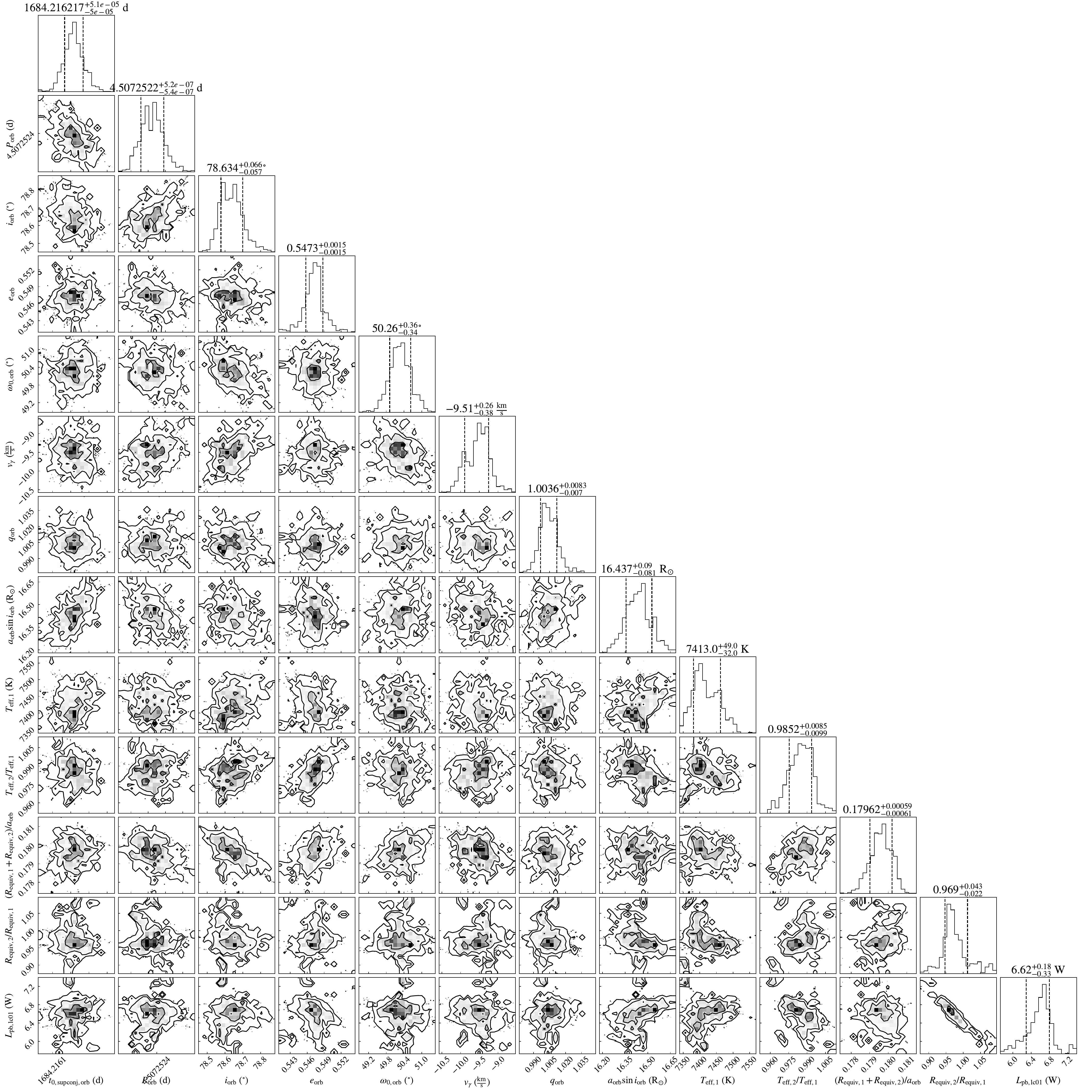}
\caption{The posterior distributions of the binary parameters optimized in the PHOEBE fits to the combined LC and RV observations.}
\label{fig7}
\end{figure}

%\newpage
\appendix
In Figure \ref{fig7}, we present the posterior distributions of the binary parameters optimized in the PHOEBE fits to the combined LC and RV observations. 
In Table \ref{tb5}, we list the combination frequencies 
that were extracted from 
the residual TESS light curve of FX UMa.
Their corresponding amplitudes, phases, and S/N are also given in Table \ref{tb5}. 

\startlongtable
\begin{deluxetable*}{lrrrrr}\label{tb5}
\tablecolumns{6}
\tablewidth{0pc}
\tablecaption{Possible combination frequencies of FX UMa}
\tablehead{\colhead{ID}   &\colhead{Frequency}   &\colhead{Amplitude}  &\colhead{Phase}   &\colhead{S/N}  &\colhead{Remark} \\
\colhead{}    &\colhead{(day$^{-1}$)}     &\colhead{(mmag)}   &\colhead{(rad/2$\pi$)}   &\colhead{}    &\colhead{}}
\startdata
$f_{12}$	 &24.01078$\pm$0.00004		&0.234$\pm$0.016		&0.012$\pm$0.011		&26.2	&$f_{2}$ + $f_{3}$ + $f_{6}$\\
$f_{13}$	 &21.57191$\pm$0.00001		&0.233$\pm$0.004		&0.729$\pm$0.004		&21.8	&$f_{1}$ + 2$f_{5}$ - 2$f_{3}$\\  
$f_{15}$	 &20.402$\pm$0.006			&0.228$\pm$0.069		&0.008$\pm$0.261		&28.8	&$f_{9}$ + 2$f_{8}$ - $f_{11}$\\
$f_{21}$	 &0.076$\pm$0.004			&0.171$\pm$0.033		&0.160$\pm$0.175		&14.5	&$f_{11}$ - $f_{15}$ - $f_{9}$\\
$f_{22}$  	 &32.77244$\pm$0.00001		&0.169$\pm$0.003		&0.299$\pm$0.004		&20.7	&2$f_{11}$ - 2$f_{4}$ - 2$f_{7}$\\
$f_{23}$	 &33.5447$\pm$0.0004		&0.163$\pm$0.007		&0.155$\pm$0.035		&20.4	&$f_{8}$ + 2$f_{18}$ - $f_{19}$\\ 
$f_{25}$	 &26.25791$\pm$0.00001		&0.145$\pm$0.003		&0.772$\pm$0.004		&18.1	&$f_{3}$ + 2$f_{18}$ - $f_{15}$\\
$f_{26}$	 &20.63576$\pm$0.00003		&0.155$\pm$0.005		&0.451$\pm$0.006		&18.4	&$2f_{19}$ - $f_{15}$ - $f_{17}$\\
$f_{27}$	 &22.459$\pm$0.006			&0.152$\pm$0.058		&0.389$\pm$0.188		&14.4	&$f_{12}$ - $f_{24}$\\
$f_{28}$	 &0.78058$\pm$0.00002		&0.118$\pm$0.004		&0.873$\pm$0.009		&10.1	&$f_{18}$ - $f_{13}$\\
$f_{29}$	 &0.139$\pm$0.009			&0.142$\pm$0.028		&0.041$\pm$0.126		&12.1	&$f_{11}$ - $f_{13}$ - $f_{24}$\\
$f_{30}$	 &28.57295$\pm$0.00002		&0.117$\pm$0.005		&0.892$\pm$0.005		&13.5	&$f_{27}$ + $f_{7}$ + $f_{9}$\\
$f_{31}$	 &25.35$\pm$0.02			&0.119$\pm$0.026		&0.489$\pm$0.188		&15.0	&$f_{18}$ + $f_{25}$ - $f_{11}$\\
$f_{33}$	 &22.22$\pm$0.07			&0.132$\pm$0.038		&0.331$\pm$0.253		&12.8	&$f_{1}$ + $f_{2}$ - $f_{16}$\\
$f_{34}$	 &38.672$\pm$0.004			&0.116$\pm$0.014		&0.092$\pm$0.115		&21.5	&$f_{1}$ + $f_{15}$ - 2$f_{3}$\\
$f_{35}$	 &29.91244$\pm$0.00002		&0.103$\pm$0.003		&0.805$\pm$0.005		&14.6	&$f_{10}$ + $f_{24}$ + $f_{5}$\\  
$f_{37}$	 &0.034$\pm$0.005			&0.118$\pm$0.013		&0.724$\pm$0.116		&10.2	&$f_{16}$ + $f_{21}$ - $f_{2}$\\ 
$f_{38}$	 &26.76590$\pm$0.00002		&0.106$\pm$0.004		&0.096$\pm$0.006		&12.4	&$f_{6}$ + 3$f_{3}$\\
$f_{39}$	 &20.32983$\pm$0.00004		&0.095$\pm$0.007		&0.704$\pm$0.015		&11.9	&$f_{38}$ - $f_{14}$ - $f_{4}$\\
$f_{44}$	 &0.21635$\pm$0.00003		&0.100$\pm$0.007		&0.234$\pm$0.011		&8.7		&$f_{21}$ + $f_{29}$\\
$f_{45}$	 &32.79152$\pm$0.00002		&0.091$\pm$0.004		&0.409$\pm$0.006		&11.1	&$f_{31}$ + $f_{43}$ + $f_{9}$\\
$f_{46}$	 &32.30360$\pm$0.00002		&0.087$\pm$0.003		&0.578$\pm$0.006		&11.1	&$f_{36}$ + $f_{45}$ - $f_{2}$\\
$f_{47}$	 &20.4420$\pm$0.0001		&0.102$\pm$0.010		&0.030$\pm$0.032		&12.7	&$f_{11}$ - $f_{37}$ - $f_{9}$\\
$f_{48}$	 &1.695$\pm$0.006			&0.085$\pm$0.020		&0.331$\pm$0.257		&7.7		&$f_{2}$ + $f_{7}$ - $f_{9}$\\
$f_{49}$	 &27.52$\pm$0.02			&0.079$\pm$0.015		&0.217$\pm$0.145		&8.2		&$f_{31}$ + $f_{7}$ - $f_{2}$\\
$f_{50}$	 &30.83245$\pm$0.00002		&0.082$\pm$0.003		&0.780$\pm$0.007		&11.9	&$f_{16}$ + $f_{2}$ + $f_{30}$\\
$f_{51}$	 &31.594$\pm$0.006			&0.089$\pm$0.034		&0.447$\pm$0.100		&13.3	&$f_{4}$ + $f_{50}$ - $f_{9}$\\
$f_{52}$	 &35.30$\pm$0.06			&0.071$\pm$0.015		&0.113$\pm$0.243		&11.1	&$f_{34}$ - $f_{2}$ - $f_{32}$\\
$f_{53}$	 &20.91$\pm$0.01			&0.078$\pm$0.013		&0.333$\pm$0.111		&8.9		&$f_{1}$ + $f_{25}$ - $f_{49}$\\
$f_{54}$	 &22.192$\pm$0.006			&0.089$\pm$0.024		&0.087$\pm$0.139		&8.7		&2$f_{1}$ - $f_{17}$\\
$f_{55}$	 &27.11080$\pm$0.00002		&0.071$\pm$0.003		&0.798$\pm$0.007		&7.7		&$f_{35}$ - $f_{2}$ - $f_{5}$\\
$f_{56}$	 &34.76155$\pm$0.00002		&0.070$\pm$0.003		&0.801$\pm$0.007		&10.4	&$f_{52}$ + $f_{9}$ - $f_{7}$\\
$f_{57}$	 &18.84781$\pm$0.00002		&0.069$\pm$0.003		&0.496$\pm$0.007		&10.8	&$f_{1}$ - $f_{7}$\\
$f_{58}$	 &20.614$\pm$0.002			&0.096$\pm$0.011		&0.303$\pm$0.086		&11.4	&$f_{18}$ + $f_{8}$ - $f_{1}$\\
$f_{59}$ 	 &0.277$\pm$0.001			&0.078$\pm$0.012		&0.951$\pm$0.101		&6.6		&2$f_{29}$\\
$f_{60}$	 &0.35$\pm$0.01			&0.079$\pm$0.019		&0.262$\pm$0.202		&6.7		&2$f_{2}$ - $f_{3}$\\ 
$f_{61}$	 &18.059$\pm$0.002			&0.067$\pm$0.009		&0.009$\pm$0.061		&11.7	&$f_{34}$ - $f_{58}$\\
$f_{62}$	 &3.9$\pm$0.1				&0.067$\pm$0.017		&0.823$\pm$0.258		&10.0	&2$f_{3}$\\
$f_{63}$	 &28.362$\pm$0.005			&0.069$\pm$0.012		&0.404$\pm$0.156		&7.5		&$f_{38}$ + $f_{4}$ - $f_{3}$\\
$f_{66}$	 &27.88186$\pm$0.00002		&0.067$\pm$0.004		&0.346$\pm$0.007		&7.2		&$f_{10}$ + $f_{3}$ - $f_{28}$\\
$f_{67}$	 &26.67935$\pm$0.00003		&0.060$\pm$0.003		&0.644$\pm$0.007		&7.1		&$f_{31}$ + $f_{40}$\\
$f_{68}$	 &28.51$\pm$0.05			&0.064$\pm$0.015		&0.017$\pm$0.142		&7.0		&$f_{11}$ + $f_{4}$ + $f_{48}$\\
$f_{69}$	 &25.156$\pm$0.002			&0.063$\pm$0.005		&0.648$\pm$0.064		&7.9		&$f_{10}$ - $f_{24}$\\
$f_{70}$	 &22.47$\pm$0.03			&0.091$\pm$0.057		&0.561$\pm$0.339		&8.6		&$f_{24}$ + $f_{53}$\\
$f_{71}$	 &33.53$\pm$0.02			&0.073$\pm$0.021		&0.390$\pm$0.266		&9.1		&$f_{17}$ + $f_{23}$ - $f_{1}$\\
$f_{72}$	 &2.34898$\pm$0.00003		&0.062$\pm$0.005		&0.499$\pm$0.010		&6.3		&$f_{11}$ - $f_{53}$\\
$f_{73}$	 &0.13$\pm$0.02			&0.076$\pm$0.031		&0.505$\pm$0.115		&6.5		&$f_{18}$ - $f_{33}$\\
$f_{75}$	 &44.09499$\pm$0.00003		&0.059$\pm$0.003		&0.717$\pm$0.008		&10.1	&$f_{11}$ + $f_{17}$ - $f_{40}$\\
$f_{76}$	 &32.58386$\pm$0.00003		&0.060$\pm$0.004		&0.041$\pm$0.009		&7.2		&$f_{2}$ + $f_{4}$ + $f_{66}$\\
$f_{77}$	 &0.60703$\pm$0.00003		&0.066$\pm$0.004		&0.770$\pm$0.009		&5.7		&$f_{1}$ - $f_{36}$ - $f_{6}$\\
$f_{79}$	 &35.266$\pm$0.008			&0.059$\pm$0.014		&0.820$\pm$0.239		&9.1		&$f_{1}$ + $f_{71}$ - $f_{8}$\\
$f_{80}$	 &24.96$\pm$0.01			&0.058$\pm$0.014		&0.801$\pm$0.177		&7.2		&$f_{68}$ - $f_{4}$\\
$f_{81}$	 &33.03311$\pm$0.00003		&0.057$\pm$0.003		&0.597$\pm$0.008		&7.0		&$f_{19}$ + $f_{5}$ - $f_{44}$\\
$f_{82}$	 &36.28$\pm$0.09			&0.055$\pm$0.019		&0.266$\pm$0.205		&10.6	&$f_{22}$ + $f_{24}$ + $f_{3}$\\
$f_{83}$	 &23.98$\pm$0.01			&0.065$\pm$0.019		&0.887$\pm$0.272		&7.5		&$f_{44}$ + $f_{7}$ + $f_{8}$\\
$f_{84}$	 &0.43802$\pm$0.00004		&0.059$\pm$0.004		&0.667$\pm$0.012		&5.1		&$f_{9}$ - $f_{72}$\\
$f_{85}$	 &40.23$\pm$0.02			&0.055$\pm$0.012		&0.135$\pm$0.160		&11.5	&$f_{1}$ + $f_{15}$ - $f_{72}$\\
$f_{86}$	 &24.03$\pm$0.01			&0.072$\pm$0.017		&0.716$\pm$0.272		&8.1		&$f_{11}$ + $f_{4}$ - $f_{9}$\\
$f_{87}$	 &41.97560$\pm$0.00003		&0.053$\pm$0.003		&0.210$\pm$0.010		&9.6		&$f_{13}$ + $f_{15}$\\
$f_{88}$	 &45.01086$\pm$0.00003		&0.052$\pm$0.003		&0.422$\pm$0.011		&8.0		&$f_{13}$ + $f_{38}$ - $f_{7}$\\
$f_{89}$	 &26.710$\pm$0.004			&0.076$\pm$0.010		&0.201$\pm$0.065		&9.0		&$f_{63}$ - $f_{5}$\\
$f_{90}$	 &30.03788$\pm$0.00003		&0.049$\pm$0.003		&0.871$\pm$0.011		&7.0		&$f_{10}$ + $f_{7}$\\
$f_{91}$	 &29.86577$\pm$0.00004		&0.049$\pm$0.003		&0.501$\pm$0.010		&6.9		&$f_{19}$ + $f_{2}$ - $f_{14}$\\
$f_{92}$	 &0.40$\pm$0.02			&0.055$\pm$0.013		&0.058$\pm$0.256		&4.7		&$f_{55}$ - $f_{10}$\\
$f_{93}$	 &21.54$\pm$0.01			&0.048$\pm$0.012		&0.178$\pm$0.218		&4.5		&$f_{13}$ - $f_{37}$\\
$f_{94}$	 &0.802$\pm$0.005			&0.052$\pm$0.008		&0.043$\pm$0.101		&4.4		&$f_{3}$ - $f_{2}$\\
$f_{95}$	 &28.541$\pm$0.004			&0.047$\pm$0.005		&0.624$\pm$0.119		&5.3		&$f_{30}$ - $f_{37}$\\
$f_{96}$	 &0.23868$\pm$0.00006		&0.051$\pm$0.006		&0.838$\pm$0.017		&4.4		&$f_{27}$ - $f_{33}$\\
$f_{98}$	 &26.740$\pm$0.002			&0.054$\pm$0.007		&0.196$\pm$0.075		&6.4		&$f_{49}$ - $f_{28}$\\
$f_{99}$	 &20.58772$\pm$0.00007		&0.051$\pm$0.005		&0.886$\pm$0.022		&6.0		&$f_{1}$ - $f_{2}$ - $f_{84}$\\
$f_{100}$	 &5.762$\pm$0.003			&0.043$\pm$0.006		&0.112$\pm$0.094		&6.3		&$f_{7}$ + $f_{9}$ - $f_{60}$\\
$f_{101}$	 &27.14271$\pm$0.00004		&0.044$\pm$0.003		&0.384$\pm$0.013		&4.7		&$f_{37}$ + $f_{55}$\\
$f_{102}$	 &48.71434$\pm$0.00004		&0.042$\pm$0.003		&0.294$\pm$0.013		&10.2	&$f_{101}$ + $f_{13}$\\
$f_{103}$	 &3.45801$\pm$0.00004		&0.042$\pm$0.003		&0.302$\pm$0.014		&5.8		&$f_{7}$ + $f_{73}$\\
\enddata
\end{deluxetable*}

\end{CJK*}
\end{document}